\documentclass[%
 reprint,
 superscriptaddress,
 amsmath,amssymb,
 aps,
]{revtex4-2}

\usepackage{graphicx}
\usepackage{dcolumn}
\usepackage{bm}
\usepackage{hyperref}
\usepackage{color}
\newcommand{\blue}[1]{{\color{blue}{#1}}}
\usepackage{bm}
\newcommand{\nc}{\newcommand}
\nc{\be}{\begin{equation}}
\nc{\ee}{\end{equation}}
\nc{\bal}{\begin{align}}
\nc{\eal}{\end{align}}
\nc{\bea}{\begin{eqnarray}}
\nc{\eea}{\end{eqnarray}}
\nc{\bean}{\begin{eqnarray*}}
\nc{\eean}{\end{eqnarray*}}
\nc{\mb}{\mbox}
\nc{\rnc}{\renewcommand}
\nc{\vk}{\mb{\bf k}}
\nc{\vp}{\mb{\bf p}}
\nc{\vn}{\mb{\bf n}}
\nc{\vq}{\mb{\bf q}}
\nc{\rr}{\mb{\bf r}}
\nc{\vz}{\hat {\mb{\bf z}}}
\nc{\vj}{\mb{\boldmath$j$}}
\nc{\vg}{\mb{\boldmath$g$}}
\nc{\x}{\mb{\boldmath$x$}}
\nc{\A}{\mb{\boldmath$A$}}
\nc{\va}{\mb{\boldmath$a$}}
\nc{\vs}{\mb{\boldmath$\sigma$}}
\nc{\vpi}{\mb{\boldmath$\pi$}}
\nc{\nab}{\nabla}
\nc{\X}{\sf x}
\nc{\kk}{{\bm{k}}}
\nc{\pp}{{\bm{p}}}
\nc{\qq}{{\bm{q}}}
\nc{\mm}{{\bm{m}}}
\nc{\upspin}{{\uparrow}}
\nc{\dspin}{{\downarrow}}
\nc{\vecq}{{\bf q}}
\nc{\veck}{{\bf k}}
\nc{\vecp}{{\bf p}}
\nc{\vecl}{{\bf l}}
\nc{\vecr}{{\bf r}}
\nc{\vecx}{{\bf x}}
\nc{\vecR}{{\bf R}}
\nc{\vecG}{{\bf G}}
\nc{\vecA}{{\bf A}}
\nc{\vecpi}{{\bf \pi}}
\nc{\vecL}{{\bf L}}
\nc{\vecK}{{\bf K}}
\nc{\im}{{\imath}}

\begin{document}

\preprint{APS/123-QED}

\title{Two-dimensional helical superconductivity and gapless superconducting edge modes in the 1T$^\prime$-WS$_2$/2H-WS$_2$ heterophase bilayer}
\author{Xuance Jiang}
\email{xuance@ucsb.edu}
\altaffiliation{Present address: Department of Chemistry and Biochemistry, University of California, Santa Barbara, Santa Barbara, California 93106, USA\\ Materials Department, University of California, Santa Barbara, Santa Barbara, California 93106, USA}
\affiliation{Center for Functional Nanomaterials, Brookhaven National Laboratory, Upton, NY 11973, USA}

\affiliation{Department of Physics and Astronomy, Stony Brook University, Stony Brook, NY 11794, USA}

\author{Jennifer Cano}

\affiliation{Department of Physics and Astronomy, Stony Brook University, Stony Brook, NY 11794, USA}

\affiliation{Center for Computational Quantum Physics, Flatiron Institute, New York, New York 10010, USA}

\author{Yuan Ping}
\affiliation{Department of Materials Science and Engineering, University of Wisconsin-Madison, WI, 53706, USA}
\affiliation{Department of Physics, University of Wisconsin-Madison, WI, 53706, USA}
\affiliation{Department of Chemistry, University of Wisconsin-Madison, WI, 53706, USA}

\author{Yafis Barlas}
\email{ybarlas@unr.edu}
\affiliation{Department of Physics, University of Nevada, Reno, 1664 N. Virginia Street, Reno, NV 89557, USA}

\author{Deyu Lu}
\email{dlu@bnl.gov}
\affiliation{Center for Functional Nanomaterials, Brookhaven National Laboratory, Upton, NY 11973, USA}

\date{\today}

\begin{abstract}
We propose a material platform comprised of transition metal dichalcogenide (TMDC) heterostructures to realize the two-dimensional (2D) helical superconductivity with an intrinsic gap. By van der Waals stacking a 2D superconductor (1T$^\prime$-WS$_2$ with inversion symmetry) on top of a 2D topological insulator (2H-WS$_2$ with mirror symmetry), the resulting TMDC bilayer exhibits Rashba superconductivity. Under an external in-plane magnetic field, the system can host finite-momentum Cooper pairing, evidenced by the divergence in the particle-particle susceptibility of a $k\cdot p$ Hamiltonian fitted to the \textit{ab initio} theory band structure.  The resulting 2D helical superconducting phase can induce superconductivity in the edge states with its spatially varying order parameter. By varying the strength of the in-plane magnetic field, we demonstrate that the helical edge state can undergo a phase transition to a one-dimensional gapless phase with narrow Fermi segments corresponding to zero-energy Bogoliubov quasi-particles. The controllable one-dimensional gapless phase serves as a clear experimental fingerprint of 2D helical superconductivity. The proposed 2D TMDC heterostructure is promising for intrinsic nonreciprocal superconducting transport and the development of Majorana-based quantum devices.
\end{abstract}
\maketitle

The emergent helical superconductivity in two-dimensional (2D) Bardeen–Cooper–Schrieffer (BCS) superconductors with strong Rashba spin-orbit coupling (SOC) and broken inversion symmetry has attracted considerable attention~\cite{gor2001superconducting,choe2019gate,yuan2022supercurrent}. Under an in-plane magnetic field, Cooper pairs on the spin-momentum-locked Fermi surfaces acquire a finite momentum \vq. Similar to the Fulde–Ferrell  state~\cite{fulde1964superconductivity}, the superconducting order parameter exhibits a uniform magnitude but a spatially varying phase, $\Delta_\mathbf{q}(\mathbf{r})=\Delta_0e^{i\mathbf{q}\cdot \mathbf{r}}$, whereas the original FF system is separated from the BCS phase under a first-order phase transition in an exchange field. The 2D helical superconductor exhibits unique ability to withstand paramagnetic limiting fields beyond the Pauli limit~\cite{gor2001superconducting}, host non-reciprocal transport in the FF-like state, i.e. the superconducting diode effect~\cite{yuan2022supercurrent,nadeem2023superconducting,hasan2024supercurrent,shaffer2025theories} and create Majorana bound states~\cite{hu2014gapless}.

Despite its importance in both basic science and quantum applications, the materials realization of 2D helical superconductivity is rare. Existing material platforms rely on proximity induced superconducting gap in the 2D electron gas~\cite{shabani2016two,suominen2017zero} or topological surface states~\cite{wang2012coexistence,xu2014momentum,hart2014induced}, but experimentally it is challenging to achieve a hard superconducting gap~\cite{tosato2023hard}. A promising alternative is to induce Rashba superconductivity directly in a 2D s-wave superconductor with interface enhanced SOC effects. 
Unlike proximity induced superconductivity~\cite{Fu2008}, the resulting superconducting gap is intrinsic and the materials properties can be tuned via interface engineering. Identifying such a material platform remains a major hurdle, because the desired properties in the heterostructure require precise control of inversion symmetry breaking and a large enough SOC effect, as well as an atomically sharp interface that is difficult to achieve due to chemical reactivity and uncontrolled nucleation~\cite{yi2022crossover}.



Experimentally, the superconducting diode effect can be used to probe helical superconducting phases~\cite{zhuang2025helical}, but it is not an unambiguous signature, as it can also arise from extrinsic factors such as junction geometry~\cite{jeon2022zero,golod2022demonstration,wu2022field}, vortex configurations~\cite{lyu2021superconducting,wu2022nonreciprocal,sundaresh2023diamagnetic}, or phase inhomogeneities~\cite{roig2024superconducting,chirolli2024diode,shaffer2024superconducting}. Moreover, phase-sensitive measurements
are experimentally demanding and often challenging to interpret. This motivates the search for alternative, robust experimental signatures of the FF-like state, particularly in material platforms where extrinsic or confounding effects can be minimized. 

In this Letter, we propose a promising material platform of 2D helical superconductors using a van der Waals (vdW) transition metal dichalcogenide (TMDC) heterophase bilayer (HPB). The HPB system consists of 1T$^\prime$-WS$_2$ stacked on 2H-WS$_2$. In the HPB, the presence of the 2H-WS$_2$ monolayer in proximity breaks the intrinsic inversion symmetry in the superconducting 1T$^\prime$-WS$_2$ monolayer and induces an enhanced Rashba SOC in 1T$^\prime$-WS$_2$ due to the interface charge transfer. The enhanced SOC creates the helical spin texture due to spin-momentum locking. This HPB can potentially be synthesized by either chemical vapor deposition, as used in the 2H/1T$^\prime$ WS$_2$ HPB~\cite{liu2024metastable} or the tear and stamp method that mechanically exfoliates TMDC monolayers and stacks them together~\cite{guo2021stacking}, thereby yielding an atomically sharp interface. 

\begin{figure} [tbh!]
\includegraphics[width=1.8 in]{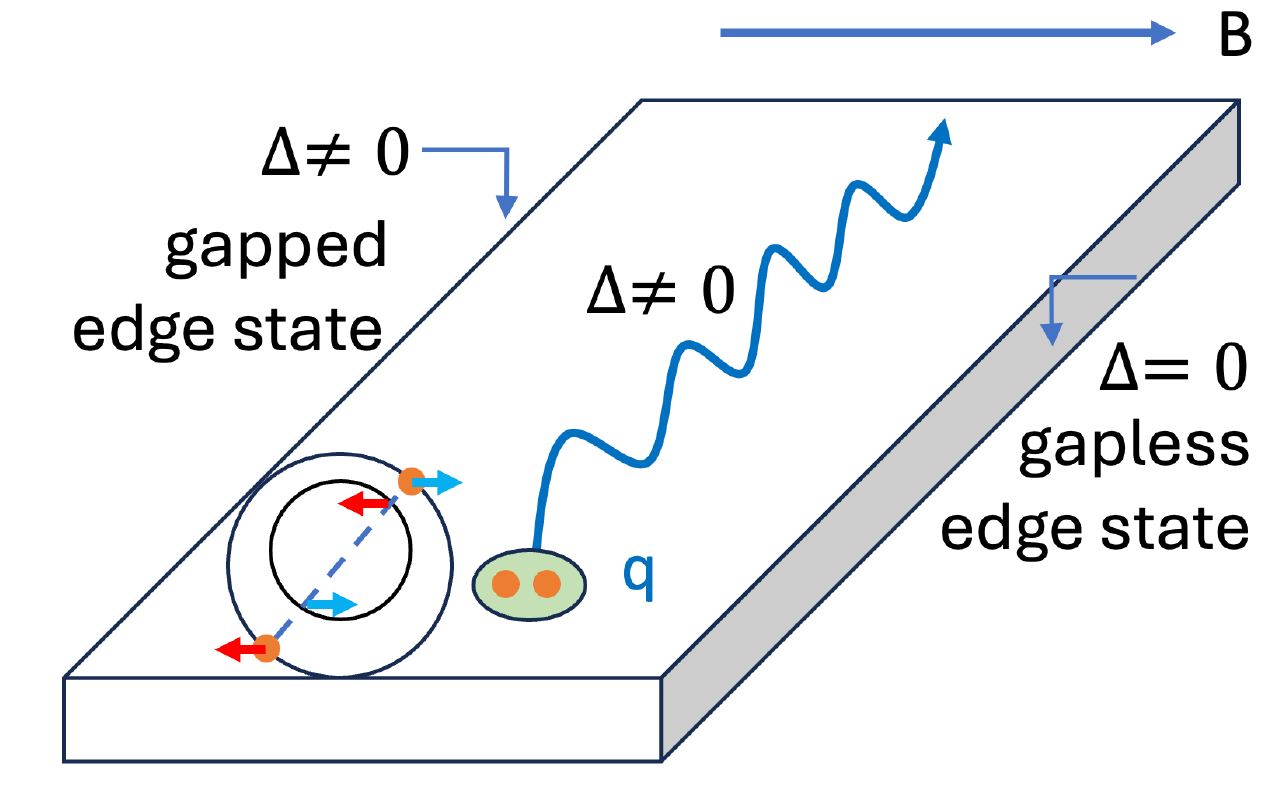}
\caption{\label{fig:cartoon} Schematic of bulk FF-like states in a 2D helical superconductor under an external field. The external field can induce the transition of the superconducting edge modes from a gapped phase to a gapless phase.} 
\end{figure}

Under an in-plane magnetic field (B), the spin-split bands are tilted, which offsets the centers of the inner and outer Fermi pockets.
As a result, Cooper pairing from nonconcentric spin helical Fermi pockets acquires a finite momentum $\mathbf{q}$~\cite{fulde1964superconductivity,larkin1965nonuniform} that gives rise to the 2D bulk FF-like state. Both $\Delta_\mathbf{q}$ and B offer the tunability over the phase diagram of the helical edge states of the HPB, which can evolve from a gapped superconducting phase to a gapless superconducting phase and finally a normal phase. This phase transition is spatially asymmetric, which allows one edge to remain gapped while the opposite edge becomes gapless, as illustrated in Fig.~\ref{fig:cartoon}. The gapless helical superconducting edge mode provides an intrinsic signature to detect the 2D FF-like state through, e.g., transport measurements. Our findings are validated by \emph{ab initio} and model Hamiltonian calculations. 

\begin{figure}[thb!]
\includegraphics[width=3.5 in]{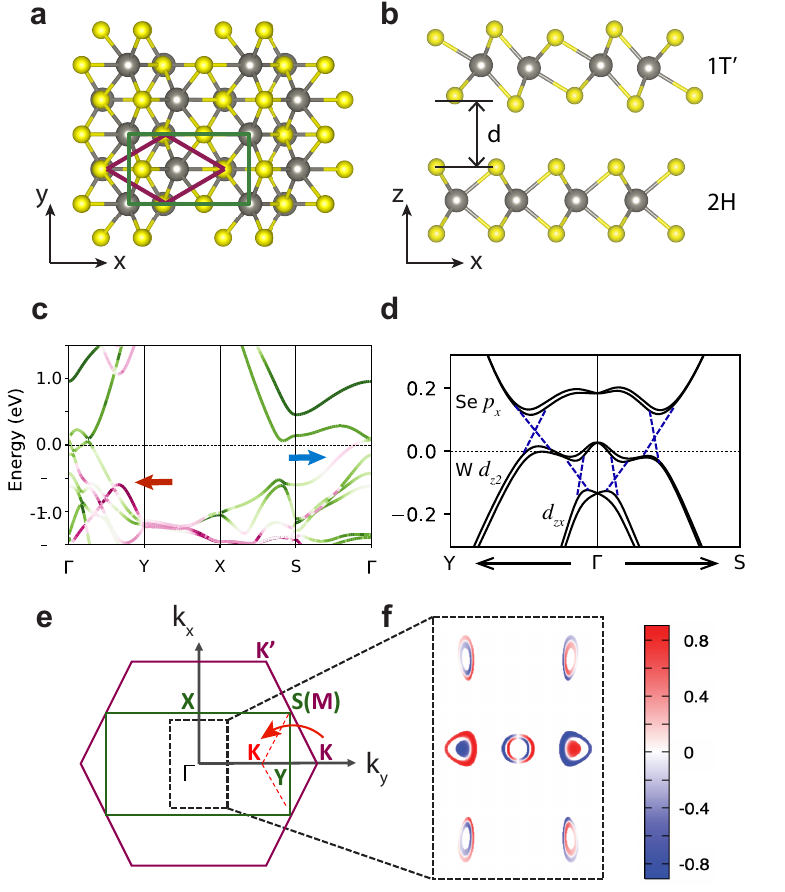}
\caption{\label{fig:bnds} (a) Top and (b) side views of the 1T$^\prime$-WS$_2$/2H-WS$_2$ heterophase bilayer. Gray and yellow spheres indicate W and S atoms, respectively. The green rectangle (purple diamond) corresponds to the unit cell of the 1T$^\prime$ (2H) phase. (c) Band structure of the heterophase bilayer, where green(purple) indicates the weight in the 1T$^\prime$(2H) layer. (d) Low energy band structure (with SOC) near $\Gamma$. Blue dashed lines indicate the bands without SOC. (e) The Brillouin zone of the 2H phase (purple) and 1T$^\prime$ phase (green). (f) Spin projection ($S_x$) of HPB Fermi pockets within the dashed rectangle. Red and blue indicate spin projection to $\pm$ x respectively.}
\end{figure}

\blue{\emph{Electronic structure of the 1T$^\prime$/2H WS$_2$ HPB}}. Bulk 1T$^\prime$-WS$_2$ is a conventional superconductor with a superconducting critical temperature ($T_c$) of 8.8 K, which decreases as the thickness of the film is reduced~\cite{lai2021metastable}. The 1T$^\prime$-WS$_2$ monolayer has inversion symmetry (see Fig.~\ref{fig:bnds}) and remains superconducting with a $T_c$ of 5.7 K~\cite{fang2019discovery,lai2021metastable}. In this study, we stack the 1T$^\prime$-WS$_2$ monolayer (2D unit cell) on top of the 2H-WS$_2$ monolayer (1$\times\sqrt{3}$ supercell). The HPB breaks the inversion symmetry in the 1T$^\prime$-WS$_2$ monolayer and the $M_z$ mirror symmetry in the 2H-WS$_2$ but preserves the $M_y$ mirror symmetry ($y\rightarrow -y$). The lattice mismatch in the HPB is compensated by compressing 1T$^\prime$-WS$_2$ and stretching 2H-WS$_2$ along the $x$ direction by 2\%, respectively. Based on simulation, a 2\% compressive strain is known to have minor effects on the superconducting properties of the 1T$^\prime$-WS$_2$ monolayer~\cite{liu2024first}.

 The HPB is studied with the density functional theory (DFT) using the Perdew-Burke-Ernzerhof (PBE) functional~\cite{Perdew1996}. Grimme's DFT-D3 method~\cite{Grimme2010} is applied to treat vdW interactions~\cite{supp}. The HPB tight-binding model is constructed using Wannier functions~\cite{Marzari2012,Pizzi2020,wanniertools}. We compare stacking configurations with different horizontal interlayer displacements~\cite{supp} and choose the one with the lowest total energy (Fig.~\ref{fig:bnds}a-b, Fig.~S1 and Table~SI~\cite{supp}), where the bottom layer S atoms in 1T$^\prime$-WS$_2$ sit on top of the W atoms in 2H-WS$_2$ at the interlayer distance ($d$) of 3.14~\AA.

The band structures of the HPB are shown in Fig.~\ref{fig:bnds}c-d. 
The hexagonal Brillouin zone of the 2H phase is folded to the rectangular Brillouin zone of the 1T$^\prime$ phase and HPB (see Fig.~\ref{fig:bnds}e). The valence band maximum (VBM) of 2H-WS$_2$ at the K point is folded to the $\Gamma$-Y line in the HPB (purple) indicated by the red arrows in Fig.~\ref{fig:bnds}c. The folded VBM is 0.54 eV below the Fermi level of the HPB. Our calculations reveal that the stacking induces a charge transfer of 0.015 e$^-$ per unit cell from the 2H phase to the 1T$^\prime$ phase. While the work function of the 1T$^\prime$ phase is comparable to the ionization energy of the 2H phase, this charge transfer creates an interface dipole potential of $\sim 0.4$ eV, which realigns the bands of the two phases (Fig.~S2~\cite{supp}). In addition, the hybridization between the 2H phase (purple) and 1T$^\prime$ phase (green) pushes the valence band of the 1T$^\prime$ bands to higher energy above the Fermi level (Fig.~\ref{fig:bnds}c, blue arrow), resulting in two hole pockets along the y axis (Fig.~\ref{fig:bnds}f). This hybridization increases the electron density of state (DOS) at the Fermi level, which enhances the $T_c$ in the HPB relative to monolayer 1T$^\prime$-WS$_2$. 

As shown in Fig.~\ref{fig:bnds}d, the low-energy bands of the WS$_2$ HPB are dominated by the S $p_x$ and W $d_{z^2}$ and $d_{zx}$ orbitals (see also Fig.~S3~\cite{supp}). More specifically, the two parabolic bands (S $p_x$ and W $d_{z^2}$) with large effective mass near $\Gamma$ originate from the 1T$^\prime$ side, consistent with previous band analysis of the monolayer 1T$^\prime$-WS$_2$~\cite{qian2014quantum}. In addition, the W $d_{zx}$ orbital from the 2H side dominates the third parabolic band with a smaller effective mass, which also crosses the Fermi level and cannot be neglected. 


%



The interfacial dipole field induces a strong Rashba SOC splitting in 1T$^\prime$-WS$_2$ under a broken inversion symmetry, a similar effect we have shown in other TMDC heterojunctions~\cite{jiang2024manipulating}. Consequently, the Fermi pockets of the HPB exhibit pronounced Rashba spin-momentum locking as shown in the spin texture in Fig.~\ref{fig:bnds}f dominated by $S_x$. Below, we show that Rashba SOC in the HPB can create a tunable FF-like state based on a low-energy $k\cdot p$ model Hamiltonian.

\blue{\emph{Low energy $k\cdot p$ model Hamiltonian}}. In the monolayer 1T$^\prime$-TMDC, a four-band $k\cdot p$ model Hamiltonian, consisting of one chalcogenide’s $p$ orbital and one TM’s $d$ orbital, plus the spin degree of freedom, can capture the low-energy dispersion and the nontrivial $\mathbb{Z}_2$ topological property~\cite{qian2014quantum}. In the HPB, the hybridization between 1T$^\prime$ and 2H phases (see Fig.~\ref{fig:bnds}d) requires a six-band  $k\cdot p$ model (one S $p$ orbital, one W $d$ orbital from the 1T$^\prime$ phase, and a hybridized 2H and 1T$^\prime$ $d$ orbital) to describe the HPB's nontrivial band topology and the spin texture of the Fermi pockets. The $k\cdot p$ Hamiltonian is given by $H_{k\cdot p}=H_{6b}+H_{ext}$, where $H_{ext}$ is the Zeeman term in the presence of a magnetic field and $H_{6b}$ is expressed as 3 by 3 blocks in the orbital space with each block a 2 by 2 matrix representing the spin degree of freedom, 
\begin{equation}
H_{6b}= \begin{pmatrix}
H^0_p + H^R_p & V_{d_1} & V_{d_2} \\
V_{d_1}^{\dagger} & H^0_{d_1} + H^R_{d_1} & D \\
V_{d_2}^{\dagger} & D & H^0_{d_2} + H^R_{d_2}
\end{pmatrix}.
\end{equation}
The diagonal blocks contain the quasiparticle energy dispersion term $H^0_n$ of orbital $n$ ($p$, $d_1$ and $d_2$) and the Rashba SOC term $H^R_n$ of the 2D system.
$H^0_n=\delta_n+\frac{\hbar^2 k_x^2}{2 m_{n_x}}+\frac{\hbar^2 k_y^2}{2 m_{n_y}}$ with $\hbar$ the Plank constant, $\delta_n$ the energy offset, $k_{i}$ ($i=x,y$) the wavevector, and $m_{n_i}$ the effective mass. The S $p$ band has a negative mass and a positive offset, whereas the two W $d$ bands have positive masses and negative offsets, indicative of the band inversion. As the stacking of 1T$^\prime$ and 2H phases breaks the inversion  symmetry of 1T$^\prime$-WS$_2$, the Rashba SOC terms are introduced in the diagonal blocks as $H^R_{n}=\lambda_n(\sigma_x k_y-\sigma_y k_x)$, where $\sigma$ are Pauli matrices and $\lambda_n$  Rashba parameters. They split the spin-degenerate bands and create spin-momentum-locked Fermi pockets. When off-diagonal SOC ($V_{n'}$ and $V_{n'}^{\dagger}$) is turned on, a topological gap is opened at the crossing between $p$ and $d$ bands, $V_{n'} = -iv_{n' x}k_x\,\mathbb{I}+v_{n' y}\sigma_x k_y$ ($n'=pd_1$ and $pd_2$ pairs), where $v_{n' i}$ are fitting parameters. Here we only consider the inversion symmetry and mirror symmetry $M_y$ allowed linear terms in $k$ in the monolayer 1T$^\prime$ TMDC~\cite{qian2014quantum}. The off-diagonal $D=t_{dd}\,\mathbb{I}$ term represents the coupling ($t_{dd}$) between the two $d$ bands. 

$H_{6b}$ was fit to the DFT band structures (Table SII~\cite{supp}) and it reproduces the low-energy band structure well within -0.04 to 0.04 eV (Fig.~S4~\cite{supp}). Compared to the spin projection of the full tight-binding model in Fig.~\ref{fig:bnds}f, the model Hamiltonian qualitatively reproduces the main Fermi pockets with the correct spin polarization and Berry curvature and the nontrivial $Z_2$ topological gap (Fig.~S5~\cite{supp}). Therefore, $H_{6b}$ provides a suitable framework to investigate Rashba superconductivity in the WS$_2$ HPB. 

\blue{\emph{Particle-particle susceptibility}}. We characterize the finite momentum pairing of the WS$_2$ HPB under an in-plane magnetic field by evaluating the static particle-particle susceptibility using the 2D band structure of $H_{k\cdot p}$~\cite{supp,yuan2021topological},
\begin{widetext}
\begin{equation}
\Pi (\qq)_{\alpha\beta} =  \int \frac{d^2 \kk} {(2 \pi)^2} \bigg[  \frac{1- n_F(\xi_{\alpha}(\kk+\qq)) - n_F(\xi_{\beta}(-\kk))}{ \xi_{\alpha}(\kk + \qq) + \xi_{\beta}(-\kk) - \im \eta} \bigg] |\langle u_{\alpha}(\kk + \qq) | u_{\beta\mathcal{T}}(-\kk) \rangle|^2,\label{eq:rf}
\end{equation}
\end{widetext}
where $\alpha, \beta = \pm$ are the two bands nearest to the Fermi level, with $u_{\alpha}$ and $\xi_{\alpha}$ their eigenvalues and eigenvectors, respectively. $u_{\beta\mathcal{T}}=i\sigma_y\mathcal{K}u_{\beta}$ is the time-reversal pair of $u_{\beta}$ with $\mathcal{K}$ the complex conjugation. Since the spin polarization is mainly along the $x$ direction, we consider only the field in the  $x$ direction, $H_{ext}=- E_z\sigma_x$, where $E_z$ is the Zeeman energy, and calculate the susceptibility as a function of $q$ along the $y$ direction. 

\begin{figure}
\begin{center}
\includegraphics[width=0.45\textwidth]{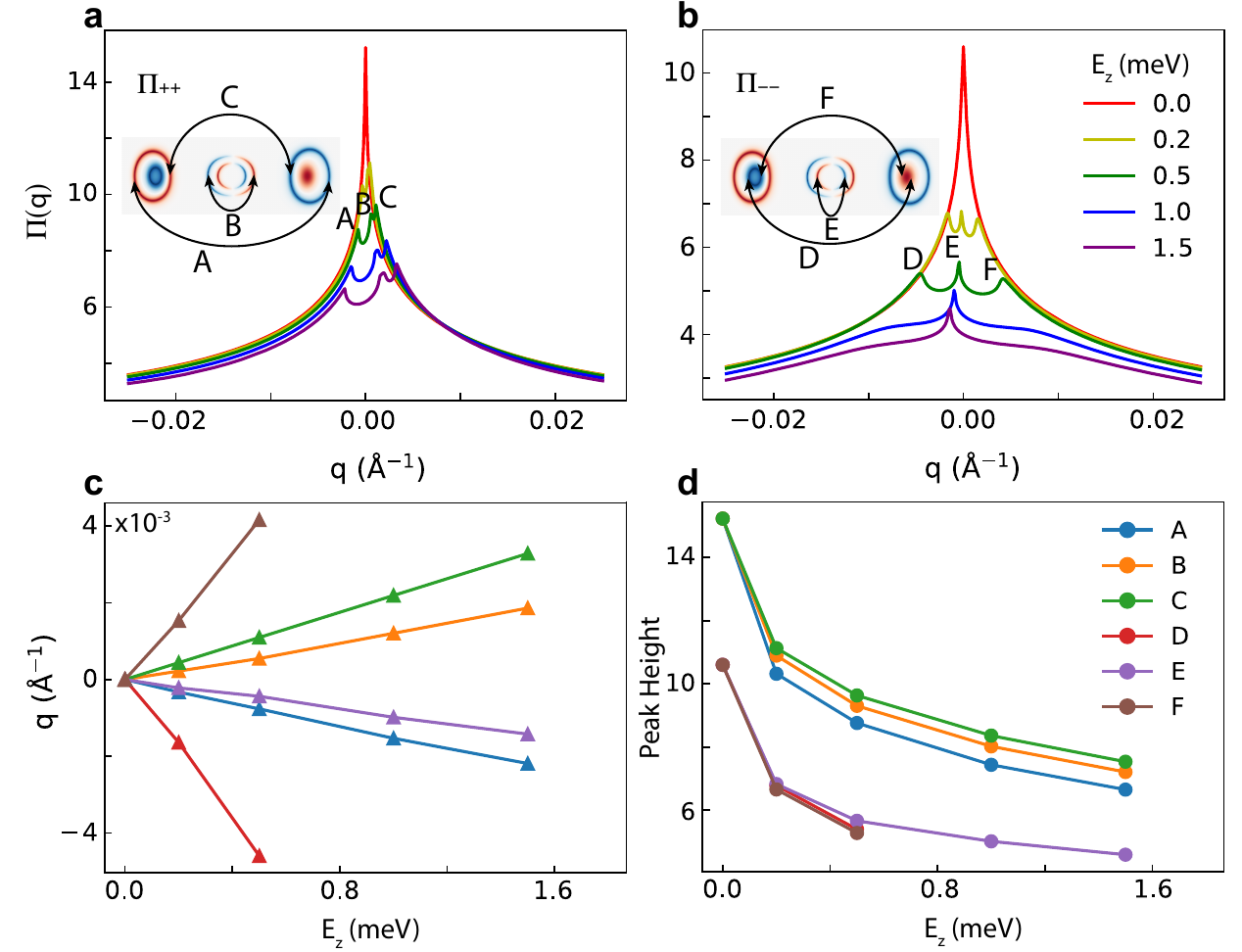}
\caption{Intraband contributions of the particle-particle susceptibility (a) $\Pi_{++}$ and (b) $\Pi_{--}$ evaluated at different Zeeman energies at temperature $\beta=1/0.0001$.  Insets show the Fermi pocket pairing at $E_z=1$ meV. (c) Position and (d) peak height of the highest logarithmic divergence in the particle-particle susceptibility as a function of $E_z$ at $\eta=10^{-7}$.} 
\label{fig:PPmodel}
\end{center}
\end{figure}

With $E_z$ ranging from 1 to 10 meV, interband susceptibilities ($\Pi_{+-}$ and $\Pi_{-+}$) are negligible compared to the intraband terms ($\Pi_{++}$ and $\Pi_{--}$).  As shown in Fig.~\ref{fig:PPmodel}, $\Pi_{++}$ is larger than $\Pi_{--}$, because of its large Fermi density of states. At zero field, the divergences of $\Pi_{++}$ and $\Pi_{--}$ happen at $q=0$, which leads to the normal s-wave Cooper pairing. However, under finite magnetic fields, the divergences are shifted to finite momentum, indicating a finite-momentum Cooper pairing. At $E_z=0.5$ meV, three peaks labeled A, B, and C correspond to the particle pairing from the outer edge of the outer left and right pockets, the outer edge of the center pocket, and the inner edge of the outer left and right pockets (see Fig.~\ref{fig:PPmodel}a-b). Similarly, three peaks labeled D, E, and F appear in $\Pi_{--}$, corresponding to the particle pairing from the outer edge of the inner left and right pockets, the inner center pocket, and the inner edge of the inner left and right pockets. While the peak positions shift linearly with increasing $E_z$ (see Fig.~\ref{fig:PPmodel}c), the peak heights decrease rapidly as $E_z$ increases (see Fig.~\ref{fig:PPmodel}d).  Therefore, divergences in $\Pi (\qq)_{\alpha\beta}$ confirm the 2D bulk FF-like state with finite momentum Cooper pairing in the WS$_2$ HPB induced by an in-plane Zeeman field.

\blue{\emph{2D FF-like state induced gapless superconducting edge states}}. 
We consider a model Hamiltonian of the helical edge state of a WS$_2$ HPB ribbon oriented along the $y$ direction (with wavevector $k$) under an in-plane magnetic field in the $x$ direction, where the bulk of the ribbon is in the helical superconducting phase. 
\begin{equation}
    H_k=H^0_k + \Delta_0 (c_{k+q/2\uparrow}^{\dagger} c_{-k+q/2\downarrow}^{\dagger} + \text{H.c.}),\label{eq:ham}
\end{equation}
where $c^{\dagger}$ and $c$ are creation and annihilation field operators. $H^0_k$ describes the left helical edge state,
\begin{equation}
H^0_k = c_k^{\dagger} \left( v_F k \sigma_x  - \mu - E_z \sigma_x \right) c_k ,
\end{equation}
where $\mu$ is the chemical potential. Rewriting Eq.~\ref{eq:ham} in the Nambu basis
$\left( 
c_{k + q/2 \uparrow},\ 
c_{k + q/2\downarrow},\ 
c^{\dagger}_{-k + q/2 \uparrow},\ 
c^{\dagger}_{-k + q/2 \downarrow} 
\right)^{\mathrm{T}}$ yields the Bogoliubov-de-Gennes (BdG) Hamiltonian,
\begin{equation}
\mathcal{H}_{BdG} = \begin{pmatrix}
H^0_{k + q/2} & i \sigma_y \Delta_0 \\
-i \sigma_y \Delta_0 & -H^{0*}_{-k + q/2}
\end{pmatrix}.
\end{equation}
The Bogoliubov energy states are
\begin{equation}
 E_k^{(\xi, \eta)} = \xi \frac{v_F q - 2E_z}{2} + \eta \sqrt{(v_F k - \xi\mu)^2 + \Delta_0^2},
\end{equation}
where $\xi, \eta =\pm 1$. On the left edge, the condition to close the superconducting gap is $|v_F q/2 - E_z| = \sqrt{(v_F k - \xi \mu)^2 + \Delta_0^2}$ at band extrema, i.e. $|v_F q/2 - E_z| = \Delta_0$. The helical state of the right edge has the opposite chirality. This leads to different gap closing conditions for two edge states as $| v_F q/2 \pm E_z| = \Delta_0$.

To estimate the superconducting temperature,  we consider the HPB similar to the 1T$^\prime$-WS$_2$ as a conventional phonon-mediated superconductor~\cite{lai2021metastable,fang2019discovery,lian2020anisotropic}. Under the condition of a weak in-plane magnetic field that does not affect the pairing, we estimate the superconducting critical temperature $T_c$ of the HPB using the Allen-Dynes formula~\cite{mcmillan1968transition,allen1975transition,carbotte1990properties}, where the isotropic Eliashberg spectral function~\cite{Giustino2007,Ponce2016} and the electron-phonon coupling strength are determined from DFT using the EPW package~\cite{Giustino2007,Ponce2016,supp}. 
The HPB phonon spectrum can largely be decomposed into 1T$^\prime$ and 2H branches with only weak hybridization near the acoustic and optical band extrema (see Fig.~S6~\cite{supp}). The calculated $T_c$ of 1.1 $K$ is in between that the bilayer (6.3 K) and monolayer 1T$^\prime$-WS$_2$ (0.4 K, see Table SIII and Figs. S7 and S8~\cite{supp}), which yields $\Delta_0\approx 0.1$ meV. 

\begin{figure}[thb!]
\includegraphics[width=3 in]{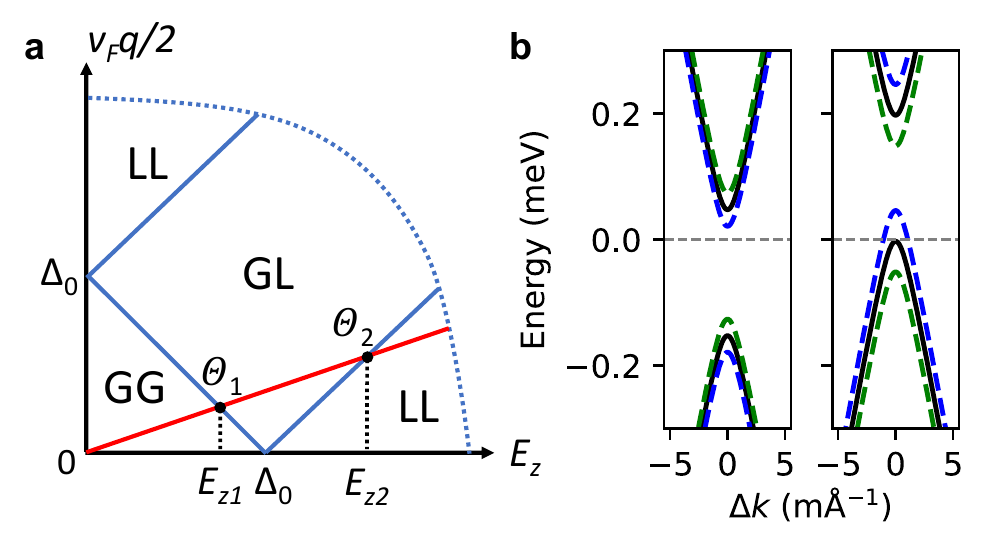}
\caption{\label{fig:phase} (a) Phase diagram of the superconducting helical edge states in the WS$_2$ HPB ribbon. GG: gapped superconducting on both edges; GL: one edge gapless; LL: both edges gapless. Red line represents $q \propto B$ with phase transition points $\mathit{\Theta_{1,2}}$ corresponding to critical Zeeman energies $E_{z1,z2}$. (b) The Bogoliubov quasiparticle band structures of two edges (left and right) at the first critical phase transition point (black solid line) near the Fermi wavevector $k_F+\Delta k$. Blue and green dashed lines indicate the 1.5$\times$ and 0.5$\times$ the critical $B$ field, corresponding to the GL and GG phase respectively. The Fermi velocity $v_F\approx 0.1$ eV$\cdot$\hbox{\AA} is used.} 
\end{figure}

The tight-binding band structure of the WS$_2$ HPB ribbons are constructed using Wannier functions~\cite{supp}, which preserves the topological helical state from the monolayer $1T^\prime$-W$S_2$ (See Fig.~S9 in~\cite{supp}). The Fermi velocity $v_F\approx 0.1$ eV$\cdot$\hbox{\AA} is extracted from the high velocity branch of the helical edge modes. 
By applying the gap closing condition for two edge states, $| v_F q/2 \pm E_z| = \Delta_0$, we obtain the phase diagram as shown in Fig.~\ref{fig:phase}a. The solid blue gap closing lines separate different superconducting edge states:  both edges gapped (GG), one edge gapped and the other gapless (GL) and both edges gapless (LL). 

In the HPB WS$_2$ ribbon, the FF-like state is induced by the in-plane magnetic field and the finite pairing momentum is proportional to the field strength. Under an increasing external field, one gapless edge state appears at the critical point $\mathit{\Theta_{1}}$ on the red line as shown in Fig.~\ref{fig:phase}a. 
The spin polarization projection $\langle S_{x} \rangle $ is about 0.3$\hbar$. If we estimate the $g$-factor of the edge state using the value of the bilayer WS$_2$ of $g_{eff}=-16$~\cite{sun2019large}, $v_Fq/2\approx 0.33 E_z$. It means that the critical Zeeman field is no longer at $E_z=\Delta_0$ without considering the effect of the finite momentum, but at a smaller value of $E_{z1}\approx0.75\Delta_0$. At this critical point, 
the gap at the left edge closes, while the gap at the right edge remains open at about 0.05 meV as shown in Fig.~\ref{fig:phase}b. The critical magnetic field is estimated at $B_1 \approx 0.1$ T, which is quite accessible in experiment. When the magnetic field is reversed, the asymmetric gap/gapless edge modes will switch sides due to the time-reversal symmetry, which provides a clear observable experimental signature. As the magnetic field increases, the system enters another phase at $E_{z2}$ with two gapless superconducting edges, if the bulk superconducting phase remains. In general, $E_{z2}$ would be larger than $\Delta_0$ as shown in Fig.~\ref{fig:phase} and we can estimate the critical magnetic field in the HPB system as $B_2  \approx 0.2$ T.

In conclusion, we propose a promising 2D transition metal dichalcogenide heterostructure, the heterophase bilayer WS$_2$, which hosts 2D helical superconductivity. The combination of 1T$^\prime$ and 2H phase TMDCs breaks the inversion symmetry and gives rise to an interface-enhanced SOC, leading to 2D helical superconductivity with an intrinsic gap. This materials design strategy can be generalized to other TMDC systems to realize 2D helical superconductors.  An in-plane magnetic field induces divergences in the pair susceptibility at finite momentum, as clear evidence of finite momentum Cooper pairing. The helical edge modes can undergo a phase transition to form gapless superconducting edge modes under a moderate external field, which provides a clear and accessible experimental signature of the FF-like state. The interplay of bulk finite-momentum pairing and boundary gapless superconductivity further opens opportunities for applications in nonreciprocal transport, spintronics, and Majorana-based quantum devices.

\vspace{0.05 in}
\section*{Acknowledgments} This research used Theory and Computation resources at the Center for Functional Nanomaterials (CFN), which is a U.S. Department of Energy Office of Science User Facility, at Brookhaven National Laboratory under Contract No. DE-SC0012704. Y.B. acknowledges support from the DOE EPSCoR program under the award DE-SC0022178. Y.P. acknowledges the funding support from the National Science Foundation under grant no. DMR-2143233. J.C. acknowledges support from the NSF under Grant No. DMR-1942447 and from the Flatiron Institute, a division of the Simons Foundation. This research used resources of the National Energy Research Scientific Computing Center (NERSC), a Department of Energy User Facility using award BES-ERCAP28324 and 32137. We thank Mark Hybertsen, Alex Levchenko, Mayank Gupta and Daniel Shaffer for helpful discussions.

\nocite{*}


%

\end{document}